# Dynamical Configurations and Bistability of Helical Nanostructures under External Torque

Arijit Ghosh,[†] Debadrita Paria,[§] Haobijam Johnson Singh,[‡] Pooyath Lekshmy Venugopalan,[§] Ambarish Ghosh[†,‡,§]

[†]Department of Electrical Communication Engineering, Indian Institute of Science, Bangalore, 560012

[‡]Department of Physics, Indian Institute of Science, Bangalore, 560012

[§]Centre for Nano Science and Engineering, Indian Institute of Science, Bangalore, 560012

**Abstract:** We study the motion of a ferromagnetic helical nanostructure under the action of a rotating magnetic field. A variety of dynamical configurations were observed that depended strongly on the direction of magnetization and the geometrical parameters, which were also confirmed by a theoretical model, based on the dynamics of a rigid body under Stokes flow. Although motion at low Reynolds numbers is typically deterministic, under certain experimental conditions, the nanostructures showed a surprising bistable behavior, such that the dynamics switched randomly between two configurations, possibly induced by thermal fluctuations. The experimental observations and the theoretical results presented in this letter are general enough to be applicable to any system of ellipsoidal symmetry under external force or torque.

Maneuvering nanoscale objects in fluidic media in a non-invasive manner can lead to various biomedical applications [1], and is pursued by researchers across several disciplines. Of particular interest [2-4] is the possibility of powering and controlling the motion of nanoscale objects with small, homogeneous magnetic fields [5], which is easy to achieve, and guaranteed to be non-invasive as well. Since motion at small length scales is dominated by viscosity, usual methods of macro scale swimming cannot lead to net locomotion; and therefore one needs to be careful in designing the shapes and symmetries of the nanoscale objects to be maneuvered. Common strategies are often based on mimicking the shapes and swimming methods of micro-organisms [6-8], such as the cork-screw motion of bacterial flagella [9] and the flexible oar-like motion of spermatozoa. This has recently been achieved by various groups using advanced nanofabrication techniques, where magnetic nanoscale objects of different shapes, such as helical, flexible rod-like [10] etc. have been maneuvered in a controllable fashion using either rotating or undulating magnetic fields. In particular, cork-screw motion is achieved in ferromagnetic helical [11,12] nanostructures by aligning the permanent magnetic moments of the helix with a rotating magnetic field, causing the nanostructure to rotate and therefore propel. Such systems have been referred to as either magnetic nanopropellers [13] or as artificial bacterial flagella [14, 15] in the literature. The purpose of this letter is to characterize the rich dynamics exhibited by these nanostructures under the action of a rotating magnetic field, and more generally , to obtain an understanding of the dynamical behavior of any cylindrically symmetric system driven by an external torque at low Reynolds numbers.

Under the action of a rotating magnetic field, an unconstrained helical object with a permanent magnetic moment, or more generally, any object with cylindrical symmetry, can rotate in a number of possible configurations. In most of the previous experiments, the magnetization of the helix was designed to be along its short axis, such that upon the action of a rotating magnetic field, the object turned about its long axis, bringing about a cork screw motion that enabled it to propel forward.  An interesting exception was the system of microhelix coils [16], where various magnetization directions wrt to a magnetic coil, including radial direction, were achieved.  As observed with the microhelix structures, and the system reported here, cork screw motion is not the only possible dynamical configuration, since the magnetic moment could remain aligned to the rotating field by turning around the short axis as well; although the later configuration requires more viscous dissipation by the helix due to the applied magnetic torque. As we show below, there are many possible dynamical configurations of a helical nanopropeller, which may or may not be such that the viscous dissipation is minimized. Although non-intuitive, this is not completely surprising from a theoretical standpoint, since there is no variational [17] principle for the Navier-Stokes equations under most general conditions; although extremum [18, 19] principles of energy dissipation have been proposed in certain cases to predict the dynamics of a solid body under Stokes flow. Interestingly, in certain range of experimental parameters, the dynamics resembled that of a bistable system, where the motion randomly switched between two possible configurations. A similar chaotic transition between rotational and oscillatory rotational motion were predicted [20] for microhelix dynamics, originating in the imbalance between magnetic and viscous torques.

***Experimental setup and observations, from tumbling to propulsion:*** The system of nanopropellers reported in the present study was fabricated using a vapor deposition technique called GLAD [21] (Glancing Angle Deposition), where thin films containing helical nanostructures (nanopropellers) could be fabricated in $SiO_2$ with a very high throughput (billion propellers / four inch wafer / evaporation). The film was sonicated in water to release the individual nanostructures, which were then laid down on a glass slide (shown in Fig. 1A), and subsequently coated with a ferromagnetic material, such as Cobalt. To investigate the dynamics of the propeller in a most general way, it was necessary to obtain various directions of magnetization with respect to the body coordinates of the nanopropeller. This was done by magnetizing the propellers along arbitrary directions (details in the supplementary information: SI).

The dynamics of the magnetized propellers were studied under the action of a rotating magnetic field, whose plane of rotation coincided with the plane of observation in the microscope. The propellers remained in solution for many hours, thus allowing them to be imaged under different field strengths in a wide range of frequencies. The observed dynamics, most generally, could be described as that of a cylinder precessing with angle $\alpha_p$ around its long axis, such that $\alpha_p = 0°$ corresponds to rotation about the long axis (referred to as "cork-screw motion"), while $\alpha_p = 90°$ corresponds to rotation about the short axis

(referred to as "tumbling").  The variation of the precession angle $\alpha_p$ as a function of frequency of the magnetic field is shown in Fig. 2 for a propeller whose direction of magnetization is somewhere between the long and the short axis, given by $\theta_m = 54°$ (angle made by the moment with the short axis). At very low frequencies *(< 8 Hz)*, the propeller "tumbles" i.e. rotates around its short axis at the frequency of the applied field (see SI, Movie SM1). Between various possible configurations (e.g. rotation around long or short axes); the propeller at low frequencies always rotated in a way that required it to overcome highest drag. Similar behavior has also been observed in artificial bacterial flagella [22], where the effect was attributed to inertial effects, and in magnetic nanorods [23], where the explanation has been based on the tendency of the system to minimize potential energy. Both these effects can be neglected in the present system, where typical Reynolds numbers (see SI) are around $10^{-4}$ (hence, minimal inertial effects) and there is no difference of potential energy between the two possible configurations. As the frequency was increased (i.e. beyond $\Omega_1$), $\alpha_p$ decreased (see SI, Movie SM2) from $90^0$ to $0^0$, at which point, the motion of the propeller started to resemble that of a cork-screw ("propulsion" region, see SI, Movie SM3). Beyond a particular frequency (referred to as the "step-out frequency" $\Omega_2$), the torque due to the applied magnetic field could not overcome the viscous drag, which caused the propeller to slow down. In a narrow frequency range around the step-out frequency, we have observed random switching between various dynamical configurations, such that the motion could not be described with a single value of $\alpha_p$, which will be discussed in a later section. For clarity, we have not shown any data for the precession angle beyond the step-out frequency in Fig. 2.

It is important to consider how the various dynamical configurations affect the speed of the propeller. In general, apart from the motion along the direction of propulsion, movement in a perpendicular direction was also observed. This is related to surface effects, which have been observed for various related systems as well [22, 24-26]. The velocity in the direction of propulsion, $v_p$, and the precession angle, $\alpha_p$, are shown in Fig. 2 as a function of frequency. In the tumbling zone, the propulsion velocity was negligibly small, while the speeds of propulsion increased with frequency in the precession and propulsion configurations. The increase continued till the step-out frequency $\Omega_2$, after which the velocity of the object slowly reduced to zero with further increase of frequency.

*Theoretical model and simulation:* To model the precessional motion of the system, we have considered an ellipsoid of dimensions *4.5 μm* and *0.9 μm* for the major and minor axes respectively (see Fig. 1B), with a direction of magnetization at angle $\theta_m$ to the short axis At any instant of time *t*, the orientation of the ellipsoids could be described by the four Euler parameters (unit quaternions), $q_0, q_1, q_2, q_3$. The rate of change of the quaternions could be related to the angular velocities ($\omega_{xBF}, \omega_{yBF}, \omega_{zBF}$) (See SI for more details). The applied torque ($\tau_{xBF}, \tau_{yBF}, \tau_{zBF}$) in the body frame, can be used to solve for the angular velocity vector $\vec{\omega}$, using $\vec{\tau} = \gamma\vec{\omega}$, assuming $\gamma = [\gamma_{xBF}\ 0\ 0, 0\ \gamma_{yBF}\ 0, 0\ 0\ \gamma_{zBF}]$ to be the friction tensor of the ellipsoid for rotational motion about the three symmetry axes. Of particular interest was the time evolution of the precession angle $\alpha_p$, given by, $cos(\alpha_p) = q_3^2 - q_2^2 - q_1^2 + q_0^2$ which settled to a constant frequency dependent value within few time periods; and was found to be independent of the initial orientation of the ellipsoid in the frequency range lower than the step-out frequency. The frequency variation of $\alpha_p$ is shown in Fig 2, which had excellent agreement with the experimental observations, with the total magnetic moment (~ $10^{-15}$ $A/m^2$) being the only adjustable parameter in the model.

To understand the propulsion speed of the propeller in a quantitative manner, the ellipsoidal model was modified to introduce an effective coupling between rotational and translational dynamics. Assuming the object to have an effective pitch, $p_{eff}$, the propulsion speed was given by $v_p = p_{eff}\Omega_{zBF}$, where $\Omega_{zBF}$ is the rotational velocity of the ellipsoid about z-axis in the body frame. We obtained excellent agreement with experimental data for $p_{eff}$ ~ 80 nm (see Fig. 2), which is somewhat lower than the geometrical pitch (See SI), and was due to hydrodynamic slip [27], also observed in rotating bacterial flagella.

*Generalized dynamics, simulation and experimental results:* The general features of the dynamics remained the same for arbitrary directions of magnetization, although the total magnetic moments of the various propellers were not exactly the same, possibly because of effects arising out of shape induced anisotropy [28,29]. Dynamics of propellers of same geometry but with different $\theta_m$ under different magnetic fields could be characterized with the two frequencies, $\Omega_1$ and $\Omega_2$, at which they ceased to tumble and propel respectively. In fig. 3, we show the variation of $\alpha_p$ with the magnetic field frequency, scaled by $\Omega_1$. Note that propellers with different $\theta_m$ under different rotating magnetic fields demonstrated universal variation with $f/\Omega_1$, except the scaled step out frequency ($\Omega_2/\Omega_1$) depended on the direction of magnetization, $\theta_m$. The dependence of $\Omega_2/\Omega_1$ on $\theta_m$ is shown in the inset of Fig. 3, along with the results from the rigid body dynamics assuming an ellipsoidal rigid body. Please note that these results and analysis are general enough to be applicable to any cylindrically symmetric body, driven by an external torque under Stokes flow. Also note that $\Omega_2/\Omega_1$ is higher for directions of magnetization along the short axis (small $\theta_m$), implying a larger propulsion region in the frequency space. To understand the dependence on the ellipsoid dimensions, we note that the two cut-off frequencies are expected to vary as $\Omega_1 \propto MB/\gamma_s$ and $\Omega_2 \propto MB/\gamma_l$. Ignoring the logarithmic terms in the formulae for the drag coefficients, we obtain $\Omega_1 \sim a^3$ and $\Omega_2 \sim ab^2$ where a and b are the semi major and semi minor axes of the ellipsoid, thus implying $\Omega_2/\Omega_1 \sim (b/a)^2$.

***Beyond step out frequency, random switching between dynamical configurations:*** Beyond the step out frequency, $\Omega_2$, the torque due to the magnetic field could not overcome the rotational drag, causing the propellers to slow down. Interestingly, in a frequency range just around $\Omega_2$, the dynamics of the propellers became unpredictable, in which the motion randomly switched between the various dynamical configurations. In Fig. 4A, we show the time evolution of the precession angle, $\alpha_p$, for a propeller driven close to its step out frequency, ~*19* Hz. The two states observed in this propeller are denoted by P (propulsion) and T (tumbling). For frequencies less than $\Omega_2$, only propulsion was observed. With further increase in frequency, the system randomly switched (see SI, Movie SM4) between propulsion and tumbling, in time scales of a few seconds, resembling the behavior of a bistable system. At even higher frequencies, the system stabilized into one dynamical configuration, here showing slow tumbling motion (see SI, Movie SM5). The corresponding propulsion velocities also showed a bistable behavior in which the speeds varied abruptly (see Fig. S2), when the propeller switched between propulsion and tumbling. To understand this behavior, we simulate the effect of the initial orientation of the propeller on the time evolution of the precession angle. The results are shown in Fig. 4B. For a frequency close to $\Omega_2$, the steady state precession angle was either 0º (propulsion) or 90º (tumbling), for 100 randomly chosen initial orientations of the propeller (see Fig. S3). The evolution of the system from one (propulsion), to two (bistable), and then to one (tumbling) configuration can be seen in Fig. 4B, where a 100-element histogram of the steady state precession angle, $\alpha_p$, has been plotted for various frequencies close to $\Omega_2$, subject to random initial orientations. The simulation results confirm the bistable nature of the nanopropeller dynamics near the step-out frequency, where the switching between the different configurations was induced by inherent thermal noise of the system. It is interesting to note that as the frequency was increased, the system went from propulsion, to bistability and then finally to tumbling, which matches well with the experimental observations.

In conclusion, we have described the complex frequency dependent dynamics of helical magnetic nanopropellers under the action of a rotating magnetic field. We have showed the direction of magnetization to be an important factor in determining the frequencies at which transitions between the various configurations occurred, and describe the relation of the precessional motion with the speed of propulsion. As far as we know, this is the first experimental attempt to study the rotation of nanorods, where the handle (in this case the magnetic moment) is at an arbitrary angle to the object (here, nanorod / nanopropeller) under rotation. The importance of magnetization direction was also observed in a related system of microhelix structures, where radially [16, 20] magnetized microcoils showed both rotational and oscillatory rotational dynamics. Under certain conditions, random switching between various dynamical configurations were observed, where the effect of the thermal energy was large enough to cause the system to switch between two possible dynamical configurations (although a chaotic back-and-forth motion has been predicted for the microcoils [20]). As far as we know, this is the first experimental observation of dynamical instability at low Reynolds numbers. The results presented here are general enough to be applicable to other ferro- and possibly paramagnetic [30] nanoscale objects with cylindrical symmetry at low Reynolds numbers. In particular, there is recent interest in composite nanostructures that have been powered through alternate means, such as catalytic motors [31,32] and magnetotactic [33] bacteria, but where the directionality of the motion is achieved through the interaction of externally applied magnetic fields with the nanostructure. It will be interesting to see if similar complexity of dynamics can also be observed in these systems, and if this can be helpful in engineering artificial nanomotors with greater functionalities. Also, the hydrodynamic interactions between the individual nanostructures should depend on their dynamical configurations, which may have an important role in the assembly of such self propelled systems. The observation of bistability in this non-equilibrium system demonstrates interesting speed fluctuations, which deserves further study. Finally, the unpredictability of the dynamics offers exciting possibilities with enhancing the control over a system of nanomotors, where identical nanostructures under the same driving force (or torque) could be made to respond, and therefore function differently.

The authors thank Winfield Hill for his help with the current amplifiers, Department of Biotechnology (DBT) and Aeronautical Development Agency (ADA-NPMASS) for funding this work, and gratefully acknowledge the usage of the facilities in Advanced Facility for Microscopy and Microanalysis (AFMM) and Micro and Nano Characterization Facility (MNCF, CeNSE) at IISc.

**Supplementary Information Available:** Text and figures describing experimental and simulation details, along with the following five movies.

This material is available free of charge via the Internet at http://prola.aps.org.

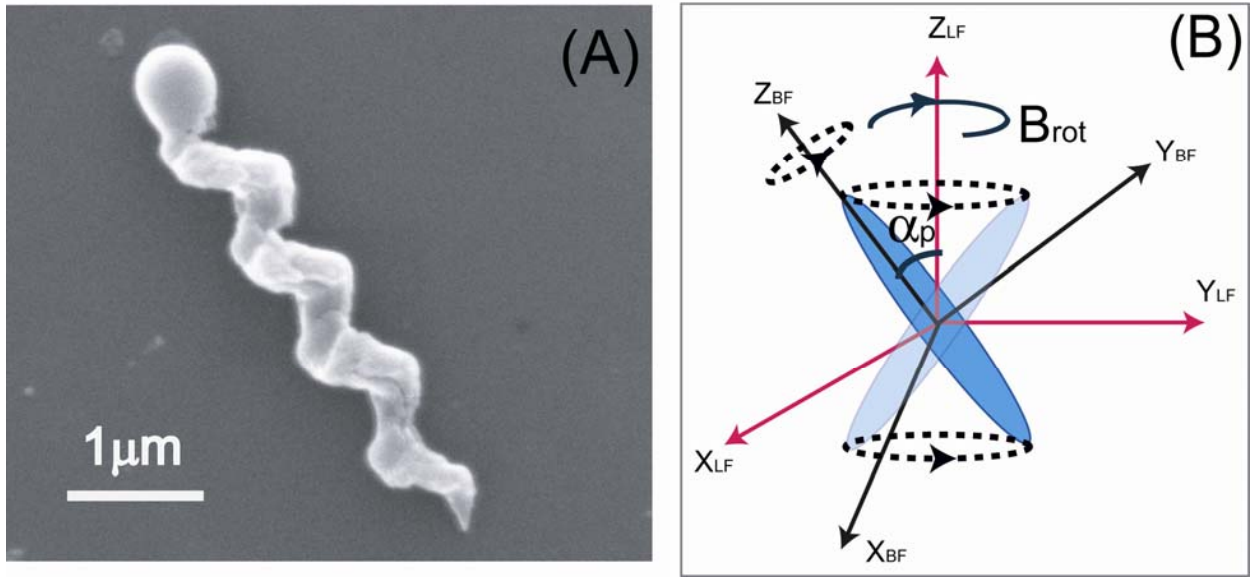

Figure 1 (color online): (A) SEM image of a single nanopropeller. (B) Schematic of the coordinate systems (BF: body frame and LF: laboratory frame) to model the precessional motion (angle $\alpha_p$) of an ellipsoid under a magnetic field, $B_{rot}$, rotating in the xy-plane.

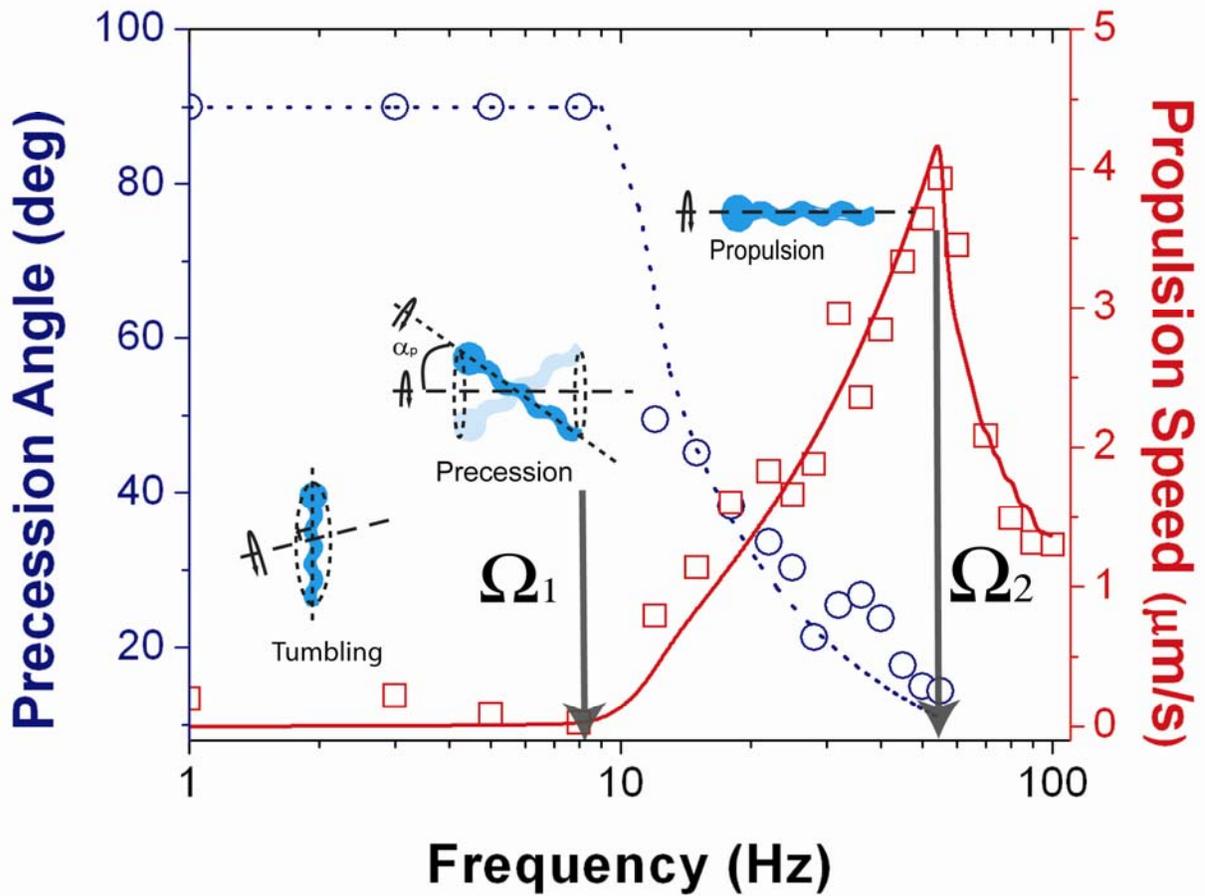

Figure 2 (color online): Experimental data for speed of propulsion (red squares) and angle of precession (blue circles) as a function of frequency for magnetic field of *20* Gauss. Inset schematics show the variety of dynamical configurations. The solid and dotted lines correspond to the propulsion speeds and precession angles of the propeller based on the theoretical model described in the text.

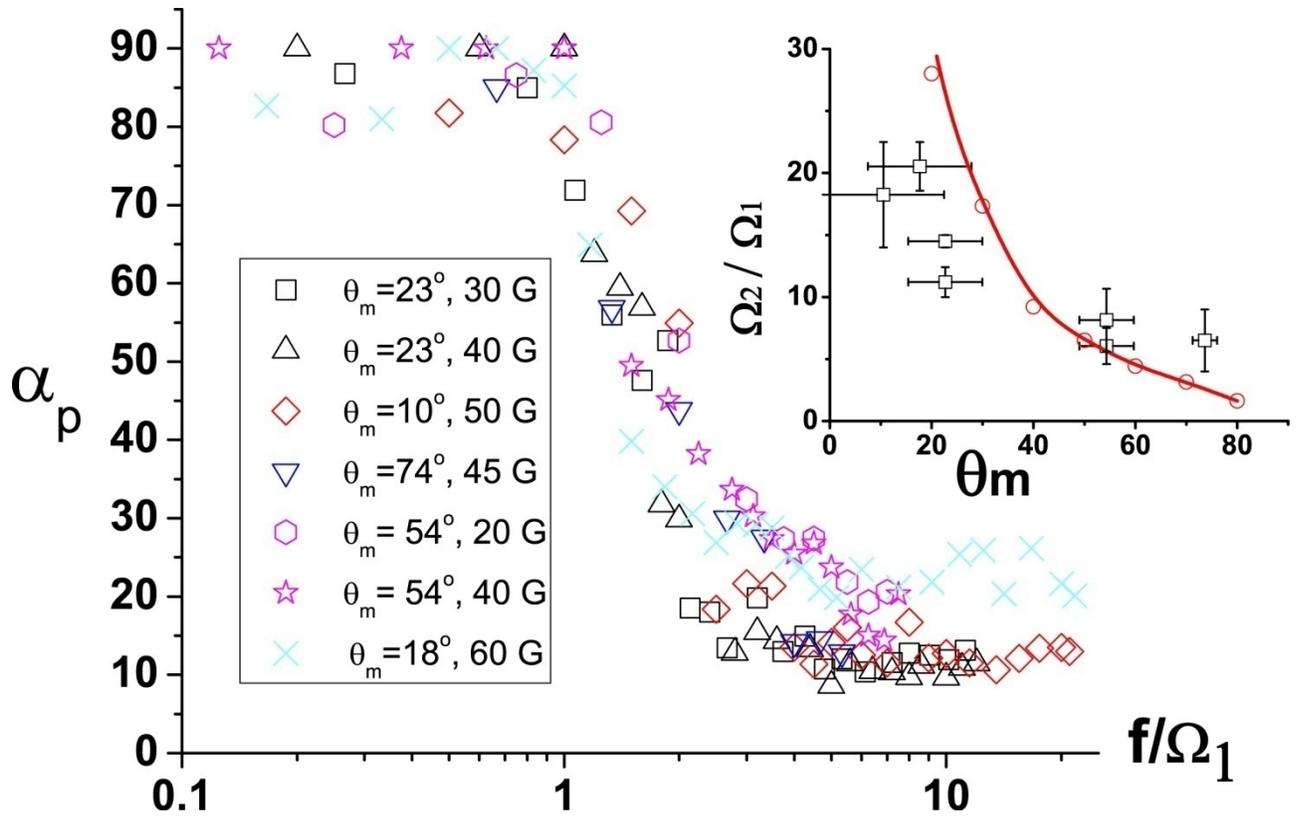

Figure 3 (color online): Variation of the precession angle as a function of the scaled frequency for propellers with various directions of magnetization under different magnetic fields. The data is plotted till $\Omega_2/\Omega_1$, which varies for the different $\theta_m$. (Inset) shows the variation of $\Omega_2/\Omega_1$ as a function of the angle of magnetization (experimental data: squares, simulation: circles and solid line).

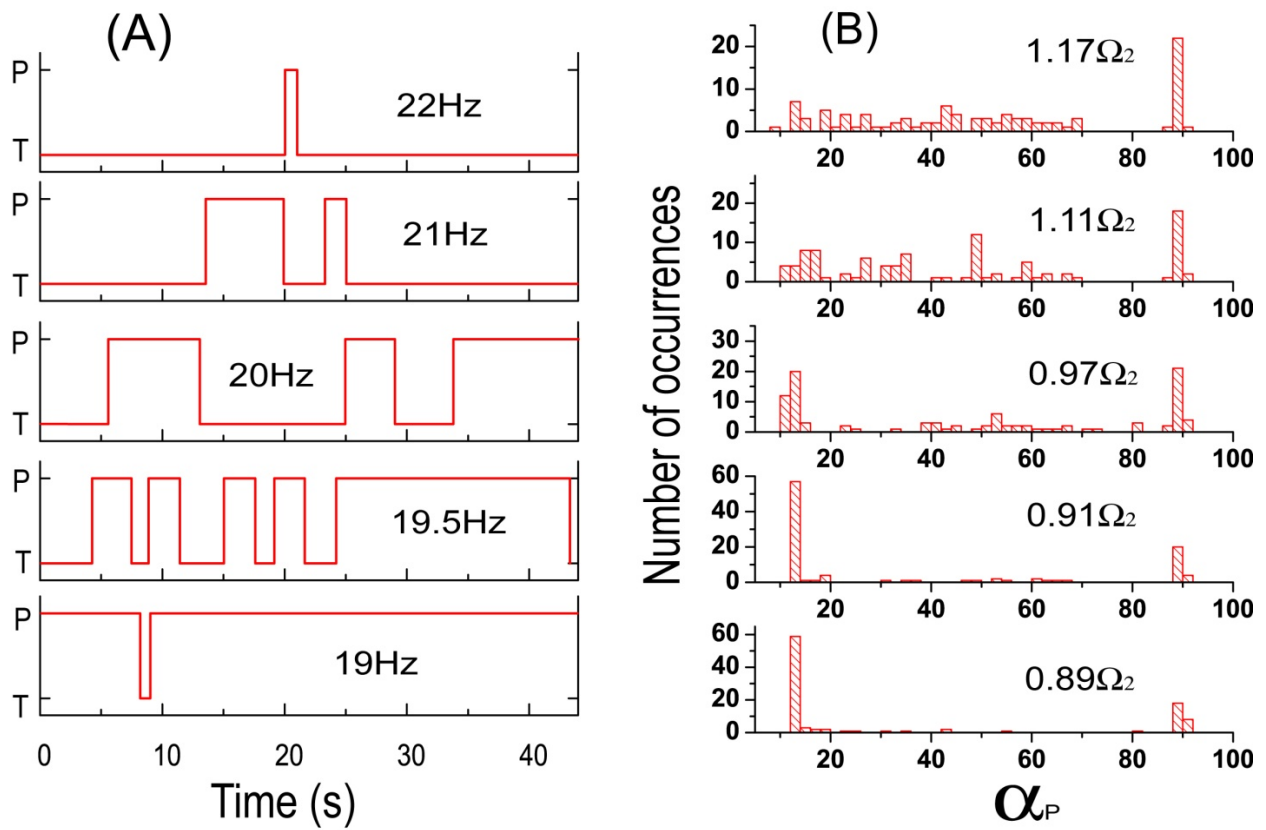

Figure 4 (color online): (A) Experimental: Time series of the dynamical configurations (denoted by "P" for propulsion and "T" for tumbling) at different frequencies. (B) Simulation: Histogram of the steady state precession angles for 100 random initial orientations of the propeller at various frequencies.